\documentclass[final,5p,times,twocolumn]{elsarticle}

\usepackage[table]{xcolor} 
\usepackage{graphicx}
\usepackage{multirow}
\usepackage{amsmath,amssymb,amsfonts}
\usepackage{amsthm}
\usepackage{mathrsfs}
\usepackage[title]{appendix}
\usepackage{textcomp}
\usepackage{manyfoot}
\usepackage{booktabs}
\usepackage{algorithm}
\usepackage{algorithmic}
\usepackage{adjustbox}
\usepackage{lscape}
\usepackage{soul}
\usepackage{pifont}
\usepackage{subcaption}
\usepackage{float}
\usepackage{tikz}
\usetikzlibrary{shapes, arrows, positioning}
\usepackage[colorlinks=true, allcolors=blue]{hyperref}

\journal{Information Fusion} 

\begin{document}

\begin{frontmatter}

\title{CrossLLM-Mamba: Multimodal State Space Fusion of LLMs for RNA Interaction Prediction}

\author[1]{Rabeya Tus Sadia}
\ead{rabeya.sadia@uky.edu}

\author[2]{Qiang Ye}
\ead{qiang.ye@uky.edu}

\author[1,3]{Qiang Cheng\corref{cor1}}
\ead{qiang.cheng@uky.edu}

\cortext[cor1]{Corresponding author}

\address[1]{Department of Computer Science, University of Kentucky, Lexington, KY, USA}
\address[2]{Department of Mathematics, University of Kentucky, Lexington, KY, USA}
\address[3]{Institute for Biomedical Informatics, University of Kentucky, Lexington, KY, USA}

\begin{abstract}Accurate prediction of RNA-associated interactions is essential for understanding cellular regulation and advancing drug discovery. While Biological Large Language Models (BioLLMs) such as ESM-2 and RiNALMo provide powerful sequence representations, existing methods rely on static fusion strategies that fail to capture the dynamic, context-dependent nature of molecular binding. We introduce CrossLLM-Mamba, a novel framework that reformulates interaction prediction as a state-space alignment problem. By leveraging bidirectional Mamba encoders, our approach enables deep ``crosstalk'' between modality-specific embeddings through hidden state propagation, modeling interactions as dynamic sequence transitions rather than static feature overlaps. The framework maintains linear computational complexity, making it scalable to high-dimensional BioLLM embeddings. We further incorporate Gaussian noise injection and Focal Loss to enhance robustness against hard-negative samples. Comprehensive experiments across three interaction categories, RNA-protein, RNA-small molecule, and RNA-RNA demonstrate that CrossLLM-Mamba achieves state-of-the-art performance. On the RPI1460 benchmark, our model attains an MCC of 0.892, surpassing the previous best by 5.2\%. For binding affinity prediction, we achieve Pearson correlations exceeding 0.95 on riboswitch and repeat RNA subtypes. These results establish state-space modeling as a powerful paradigm for multi-modal biological interaction prediction.
\end{abstract}

\begin{keyword}
Large Language Models \sep State Space Models \sep Mamba Architecture \sep RNA Interaction Prediction
\end{keyword}
\end{frontmatter}

\section{Introduction}
Ribonucleic acids (RNAs) are fundamental drivers of cellular complexity, orchestrating gene expression, splicing, and translation through intricate interactions with proteins, small molecules, and other RNAs. The dysregulation of these interactions is a hallmark of numerous pathologies, making the accurate prediction of RNA-associated complexes a critical task in modern drug discovery and systems biology. 

The recent emergence of Biological Large Language Models (BioLLMs), such as ESM-2 \cite{lin2022language} for proteins and RiNALMo \cite{penic2025rinalmo} for RNA, has provided powerful tools for sequence encoding by capturing rich semantic information within high-dimensional latent spaces. However, a significant bottleneck remains in how to effectively \textit{fuse} these multimodal representations to predict interactions. Existing computational methods predominantly rely on static fusion paradigms \cite{pan2019recent}, where biological embeddings are treated as fixed feature vectors combined via concatenation, element-wise averaging, or shallow gating mechanisms. 

While effective for simple pattern matching, static fusion fails to capture the dynamic, non-linear structural dependencies that govern molecular binding. It treats the interaction as a mere overlap of features rather than a complex biological ``dialogue'' where the conformation and state of one molecule inherently condition the binding potential of the other. Furthermore, standard training objectives often overlook the severe class imbalance and ``hard-negative'' nature of biological interaction datasets, leading to models that generalize poorly to novel sequences.

To bridge this gap, we introduce \textbf{CrossLLM-Mamba}, a novel framework that reformulates biological interaction prediction as a State-Space Modeling (SSM) alignment problem. Drawing inspiration from the efficiency of Selective State Space Models (Mamba) \cite{gu2023mamba}, we hypothesize that the interaction between two biological entities can be modeled as a sequence transition. By stacking the representations of interacting partners, we allow the ``hidden state'' of one modality to flow into and modulate the representation of the other, capturing the contextual crosstalk that static gating misses.

The main contributions of this work are summarized as follows:
\begin{itemize}
    \item \textbf{State-Space Interaction Modeling:} We propose a novel paradigm that treats biological interaction as a state-transition process rather than a static feature fusion. By utilizing the Mamba architecture, we enable deep ``crosstalk'' between LLM-generated embeddings through hidden state propagation.
    \item \textbf{Linear Complexity for High-Dimensional LLMs:} Unlike Transformer-based cross-attention which scales quadratically, our Mamba-based mixer maintains linear complexity. This allows for the efficient processing of high-dimensional embeddings from state-of-the-art BioLLMs like ESM-2 and RiNALMo without excessive computational overhead.
    \item \textbf{Multi-modal Flexibility:} The CrossLLM-Mamba framework is designed to be modality-agnostic. We demonstrate its versatility by predicting interactions across three distinct categories: RNA-protein, RNA-RNA, and RNA-small molecule interactions.
    \item \textbf{Robust Training Regime:} To address the noise and class imbalance inherent in biological data, we integrate Gaussian noise injection for data augmentation and Focal Loss \cite{lin2017focal} for training. This combination enhances the model's ability to distinguish hard-negative samples and improves generalization to unseen sequences.
\end{itemize}

\section{Related Work}
The prediction of RNA-associated interactions has evolved from traditional sequence-based statistical methods to sophisticated deep learning and large-scale language model (LLM) frameworks. We review the progression of methods across different interaction types, the emergence of biological foundation models, and the growing application of state-space models in computational biology.

\subsection{RNA-Protein Interaction Prediction}
Early computational approaches for RNA-protein interaction (RPI) relied primarily on handcrafted features and classical machine learning. RPISeq-RF \cite{muppirala2011predicting} utilized k-mer compositions with Random Forest classifiers, establishing sequence-based prediction as a viable approach. The field shifted toward deep learning with IPMiner \cite{pan2016ipminer}, which employed stacked autoencoders to automatically extract high-level features from RNA and protein sequences. Subsequent architectures introduced increasing complexity: RPITER \cite{peng2019rpiter} proposed hierarchical deep learning frameworks for lncRNA-protein pairs, while CFRP \cite{dai2019construction} focused on constructing complex handcrafted features combined with neural networks. More recently, LPI-CSFFR \cite{huang2022lpi} demonstrated the benefits of serial fusion and feature reuse for capturing long non-coding RNA dependencies, and RNAincoder \cite{wang2023rnaincoder} introduced convolutional autoencoders specifically designed for RNA-associated interactions.

\subsection{RNA-RNA Interaction Prediction}
Predicting interactions between RNA molecules, particularly miRNA-target and miRNA-lncRNA interactions, presents unique challenges due to the structural complementarity and thermodynamic stability requirements governing RNA duplexes. Traditional methods such as RNAhybrid \cite{kruger2006rnahybrid} and IntaRNA \cite{mann2017intarna} employed thermodynamic modeling to predict binding sites based on minimum free energy calculations. Machine learning approaches subsequently emerged, with PmliPred \cite{kang2020pmlipred} introducing hybrid models combining multiple classifiers with fuzzy decision rules for plant miRNA-lncRNA prediction. CORAIN \cite{wang2023corain} addressed the challenge of cross-species generalization by developing task-specific encoding algorithms based on convolutional autoencoders. However, these methods typically focus on sequence complementarity patterns and may miss higher-order dependencies captured by modern language models.

\subsection{RNA-Small Molecule Interaction Prediction}
The prediction of RNA-small molecule binding is critical for RNA-targeted drug discovery, yet remains less explored than protein-ligand interaction prediction. Early approaches adapted protein-ligand methods, using molecular fingerprints and traditional machine learning \cite{warner2018principles}. RSAPred \cite{krishnan2024reliable} aggregated experimental binding data and employed machine learning for affinity prediction across RNA subtypes. RLaffinity \cite{sun2024rlaffinity} advanced the field by combining contrastive pre-training with 3D convolutional neural networks to capture spatial binding information. Despite these advances, most methods treat RNA and small molecule representations independently before late-stage fusion, potentially missing the dynamic interplay between binding partners.

\subsection{Biological Large Language Models}
The advent of foundation models has revolutionized biological sequence encoding. For proteins, ESM-2 \cite{lin2022language} leveraged transformer architectures trained on millions of evolutionary sequences to capture deep structural and functional grammars within high-dimensional latent spaces. The model's attention patterns have been shown to correlate with protein contact maps, suggesting that sequence models implicitly learn structural information. For RNA, RiNALMo \cite{penic2025rinalmo} emerged as a large-scale foundation model capable of generalizing across structure prediction tasks, capturing the unique secondary structure propensities and evolutionary constraints of RNA sequences. In the small molecule domain, MoleBERT \cite{xia2023mole} and related graph neural network approaches encode chemical topologies through pre-training on molecular graphs.

While these BioLLMs provide superior embeddings, the challenge has shifted toward effectively fusing multi-modal representations. Recent frameworks like BioLLMNet \cite{abir2025biollmnet} explored specialized transformation networks for cross-LLM embedding fusion. However, such approaches largely rely on static fusion mechanisms, concatenation, element-wise operations, or shallow gating that may not fully capture the dynamic, context-dependent crosstalk between interacting molecules.

\subsection{Multi-Modal Fusion Strategies}
The fusion of heterogeneous representations is a fundamental challenge in multimodal learning. In the vision-language domain, cross-attention mechanisms \cite{vaswani2017attention, lu2019vilbert} have proven effective for modeling interactions between modalities, allowing one modality to attend to relevant features in another. However, cross-attention scales quadratically with sequence length, presenting computational challenges for high-dimensional biological embeddings.

Alternative fusion strategies include bilinear pooling \cite{fukui2016multimodal}, which captures pairwise feature interactions but suffers from parameter explosion, and tensor fusion networks \cite{zadeh2017tensor}, which model uni-modal, bi-modal, and tri-modal interactions explicitly. Gated fusion mechanisms \cite{arevalo2017gated} learn to weight modality contributions dynamically but still treat the fusion as a single-step aggregation rather than a continuous process.

In computational biology, most interaction prediction methods employ late fusion, where modality-specific encoders produce fixed representations that are subsequently combined \cite{pan2019recent}. This paradigm treats interaction as a static overlap of features rather than a dynamic process where the conformation and state of one molecule conditions the binding potential of the other. Our work addresses this limitation by reformulating fusion as a sequential state transition, allowing continuous information flow between modalities.

\subsection{State Space Models in Computational Biology}
State Space Models (SSMs), particularly the Mamba architecture \cite{gu2023mamba}, have emerged as powerful alternatives to Transformers due to their linear scaling complexity and selective memory mechanisms. The selective state-space formulation allows the model to dynamically filter information based on input content, making it well-suited for processing long biological sequences.

In genomics, Mamba-based architectures have shown promise for DNA sequence modeling. Caduceus \cite{schiff2024caduceus} introduced a bidirectional Mamba architecture for DNA language modeling, demonstrating that SSMs can capture long-range dependencies in genomic sequences while maintaining computational efficiency. Similarly, applications in protein structure prediction and variant effect prediction have begun exploring SSM-based alternatives to attention mechanisms.

However, the application of Mamba as a multi-modal alignment tool for biological interaction prediction remains largely unexplored. Existing SSM applications in biology focus primarily on single-modality sequence modeling rather than cross-modal fusion. Our work bridges this gap by reformulating interaction prediction as a bidirectional state-space alignment problem, leveraging Mamba's recurrent hidden state to enable deep crosstalk between LLM-generated embeddings from different biological modalities.

\subsection{Summary and Positioning}
Existing RNA interaction prediction methods face three key limitations: (1) static fusion strategies that fail to model dynamic molecular crosstalk, (2) quadratic complexity of cross-attention mechanisms that limits scalability to high-dimensional BioLLM embeddings, and (3) limited generalization across interaction types and species. CrossLLM-Mamba addresses these challenges through a unified state-space alignment framework that enables continuous information flow between modalities with linear complexity, while robust training strategies enhance generalization to unseen sequences.

\begin{figure*}[htbp]
    \centering
    \includegraphics[width=\linewidth]{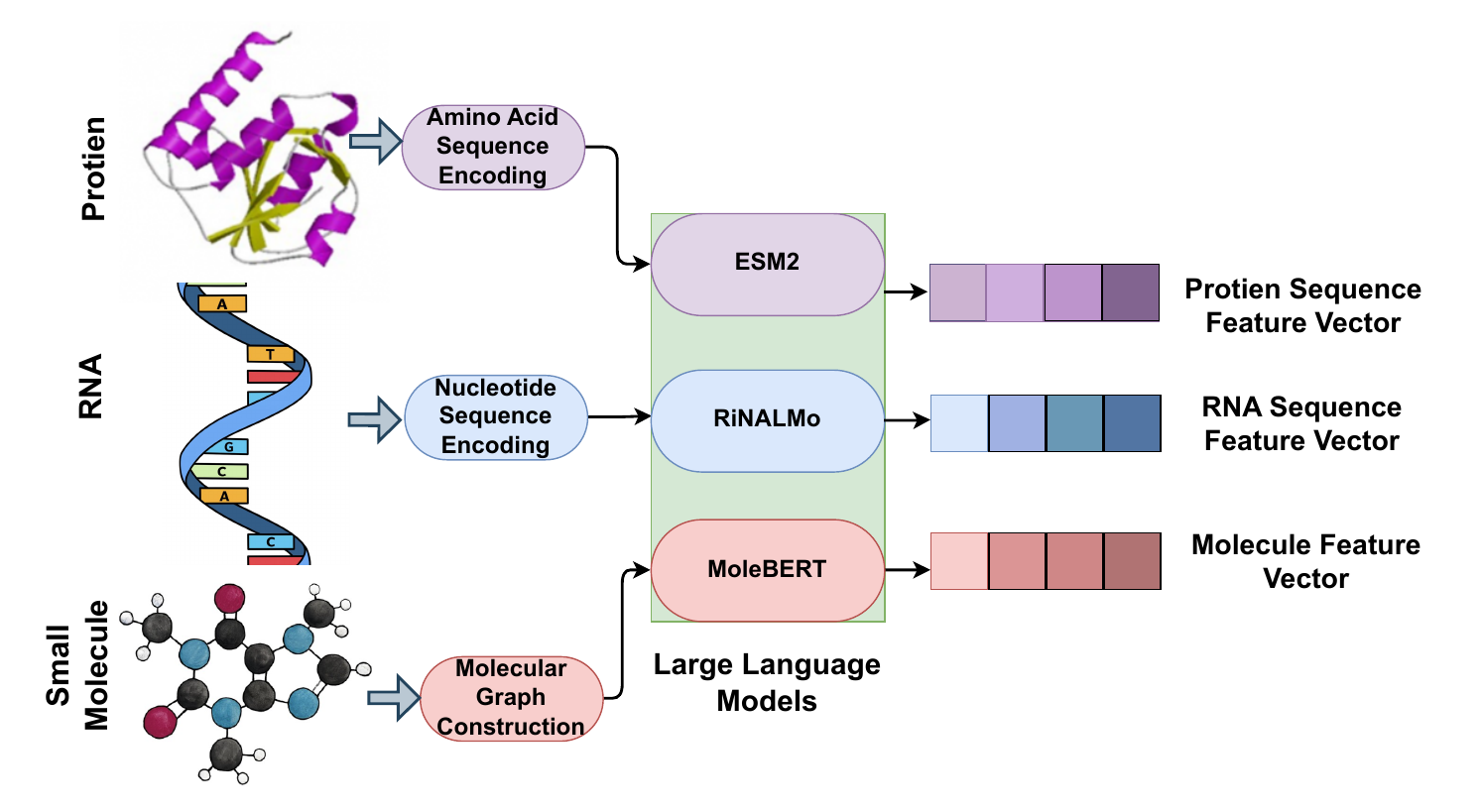}
    \caption{\textbf{Sequence Embedding and Feature Extraction Pipeline.} 
    The proposed framework utilizes specialized pre-trained large language models to encode biological entities into high-dimensional feature vectors. (Top) Protein amino acid sequences are encoded using \textbf{ESM2}. (Middle) RNA nucleotide sequences are processed via \textbf{RiNALMo}. (Bottom) Small molecule graphs are constructed and encoded using \textbf{MoleBERT}. These distinct embedding streams serve as the initial input features for the downstream dual-path Mamba architecture.}
    \label{fig:embedding_process}
\end{figure*}

\section{Overview of CrossLLM-Mamba}
\begin{figure*}[htbp]
    \centering
    \includegraphics[width=\linewidth]{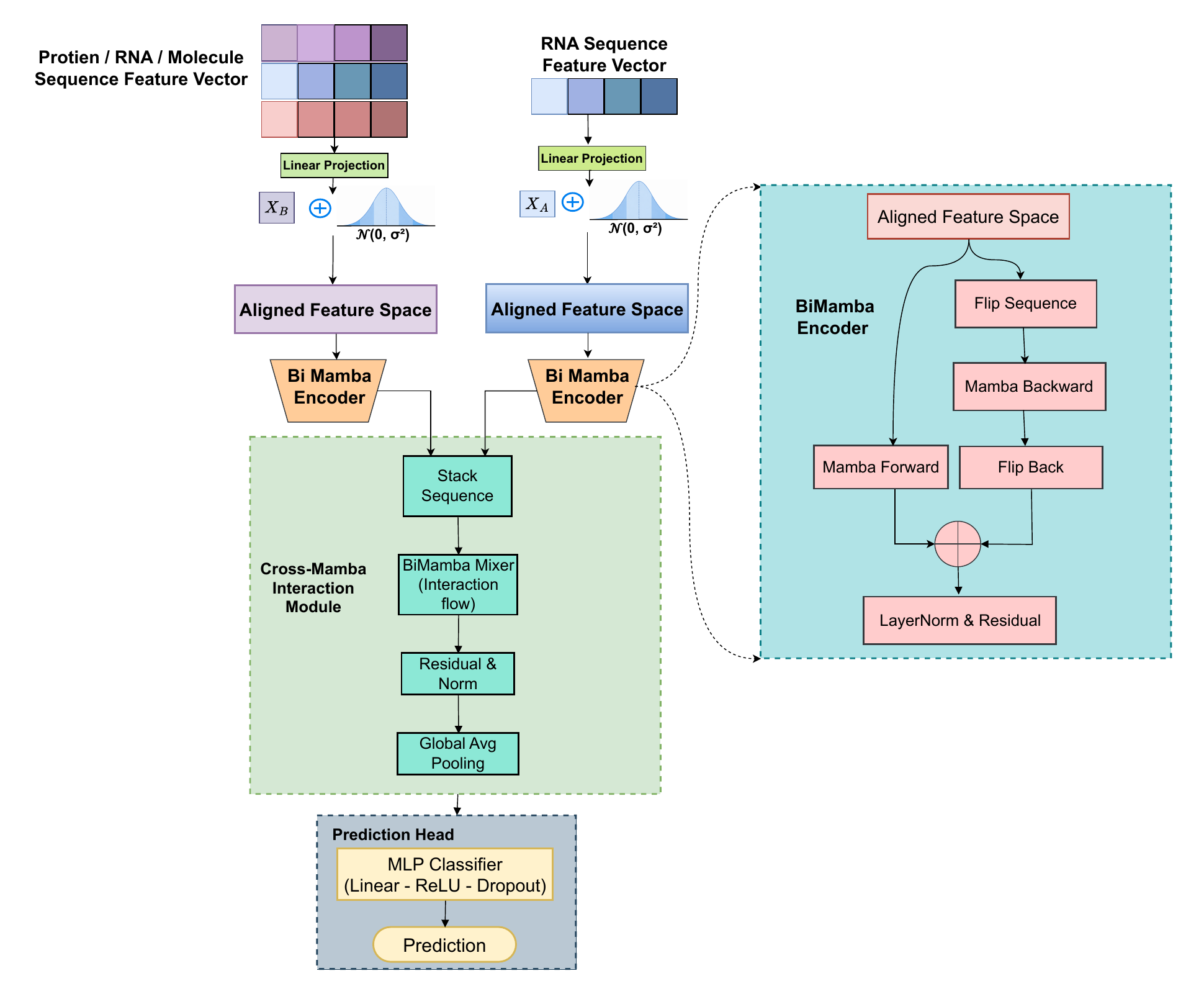}
    \caption{\textbf{The CrossLLM-Mamba Model Architecture.} 
    The framework processes multi-modal inputs (Protein, RNA, or Molecule feature vectors) through a dual-path pipeline. 
    First, feature vectors are projected and aligned using a linear transformation with Gaussian noise injection ($\mathcal{N}(0, \sigma^2)$) to enhance robustness. 
    These aligned features are encoded by parallel \textbf{BiMamba Encoders} (detailed in the right panel), which capture bidirectional sequential dependencies. 
    The encoded representations are fused in the \textbf{Cross-Mamba Interaction Module} via sequence stacking and a BiMamba Mixer to explicitly model interaction flows. 
    Finally, global average pooling aggregates the features for the MLP prediction head to output the interaction probability.}
    \label{fig:crossllm_mamba_arch}
\end{figure*}

We present CrossLLM-Mamba, a unified deep learning framework designed to predict RNA-associated interactions by reformulating the task as a state-space modeling problem. CrossLLM-Mamba first encodes biological sequences from each modality using pretrained language models specifically designed for RNA, proteins, or small molecules. Unlike prior approaches that rely on complex transformation networks and static gating mechanisms, we employ a robust noise-injected projection followed by a Cross-Mamba Interaction Module.\\
The initial stage of our framework involves a multi-modal embedding pipeline designed to convert raw biological data into semantically rich feature vectors (Fig. \ref{fig:embedding_process}). We utilize specialized foundation models as frozen feature extractors: ESM-2 for protein amino acid sequences, RiNALMo for RNA nucleotides, and MoleBERT for small molecule SMILES strings. This strategy ensures that the unique structural and evolutionary grammar of each modality is preserved before the sequences are projected into a shared latent space for interaction modeling.
Central to our approach is the introduction of a \textit{Bidirectional State Space Model (BiMamba)} that treats the interaction between two biological entities as a dynamic sequence transition. This allows the model to capture global, non-causal dependencies within frozen embeddings and enables "state mixing" where the latent representation of one molecule dynamically informs the other. The full pipeline, including the BiMamba encoders and the Cross-Mamba fusion module, is trained end-to-end. A schematic diagram of the architecture is shown in Figure \ref{fig:crossllm_mamba_arch}. The details of each module are described below:

\subsection{Language Model Embeddings for Biological Modalities}
We begin by independently encoding the input sequences of RNAs, proteins, and small molecules using pretrained language models specialized for each type of biological modality. A schematic representation of the extraction of the embeddings is shown in Figure \ref{fig:embedding_process}.

\subsubsection{RNA Language Model Embeddings}
RNA sequences, composed of nucleotides adenine (A), uracil (U), cytosine (C), and guanine (G), are embedded using \textbf{RiNALMo}, a large-scale RNA foundation model. Given an input RNA sequence $\mathcal{S}=\{s_{1},s_{2},...,s_{n}\}$, RiNALMo produces a global feature embedding of dimension 1280. These embeddings capture structural and contextual dependencies across the RNA sequence, providing a rich initial representation for downstream interaction modeling.

\subsubsection{Protein Language Model Embeddings}
Protein sequences $\mathcal{P}=\{p_{1},p_{2},....,p_{m}\}$ are encoded using \textbf{ESM-2}, a transformer-based model trained on millions of protein sequences. The model tokenizes the input amino acid sequence and outputs contextualized embeddings of dimension 1024. The representation corresponding to the token is utilized as a global summary of the protein, serving as the primary input for protein-RNA interaction tasks.

\subsubsection{Small Molecule Language Model Embeddings}
For small molecules, represented by SMILES strings $S$, we use \textbf{MoleBERT} \cite{xia2023mole} to generate molecular embeddings. MoleBERT parses the SMILES string into a molecular graph $G=(V,E)$, where each atom $v_{i}\in V$ is encoded into a latent vector $z_i$. These vectors are processed to capture both atomic and topological features. The final molecular representation is obtained through graph-level pooling, resulting in a feature vector of dimension 768.

We select RiNALMo, ESM-2, and MoleBERT as backbone language models because they represent the state-of-the-art in their respective domains. Their ability to capture rich, modality-specific semantics provides a robust foundation for our unified framework.

\subsection{Robust Feature Alignment via Noise Injection}
The raw embeddings generated from these foundation models reside in vector spaces of varying dimensions (RNA: 1280, Protein: 1024, Molecule: 768). To enable interaction modeling, we first project these embeddings into a shared latent space of dimension $D$.

Unlike previous methods that rely on deep transformation networks, we employ a linear projection augmented with \textbf{Gaussian Noise Injection}. Let $E_A$ and $E_B$ be the source embeddings. We compute:
\begin{equation}
    X_A = \mathbf{W}_A E_A + \mathbf{b}_A + \mathcal{N}(0, \sigma^2)
\end{equation}
\begin{equation}
    X_B = \mathbf{W}_B E_B + \mathbf{b}_B + \mathcal{N}(0, \sigma^2)
\end{equation}
where $\mathcal{N}(0, \sigma^2)$ represents additive Gaussian noise applied during training (with $\sigma=0.02$). This stochastic regularization forces the downstream Mamba models to learn robust structural dependencies rather than overfitting to specific artifacts in the high-dimensional LLM latent space, significantly mitigating the "hard-negative" problem common in biological datasets.

\subsection{Bidirectional Mamba Encoder (BiMamba)}
Standard State Space Models (SSMs) \cite{gu2023mamba} like Mamba are inherently causal, processing sequences from left to right. However, the global embedding of a molecule does not possess a strict temporal order. To fully capture the context within the projected features, we utilize a Bidirectional Mamba (BiMamba) block.

We treat each projected feature vector $X$ as a sequence length. The BiMamba block processes this sequence in both forward and reverse directions:
\begin{align}
    H_{\text{fwd}} &= \text{Mamba}(X) \\
    H_{\text{bwd}} &= \text{Flip}(\text{Mamba}(\text{Flip}(X)))
\end{align}
The forward and backward states are concatenated and projected back to dimension $D$, followed by a residual connection and layer normalization:
\begin{equation}
    X_{\text{enc}} = \text{LayerNorm}(\text{Linear}([H_{\text{fwd}} \mathbin{\|} H_{\text{bwd}}]) + X)
\end{equation}
This ensures that the encoder captures non-causal dependencies and global context effectively.

\subsection{Cross-Mamba Interaction Fusion}
The core innovation of CrossLLM-Mamba is the Cross-Mamba Interaction Module. Instead of static element-wise gating or simple concatenation, we model the interaction as a sequential state transition.

We construct a unified interaction sequence $S$ by stacking the encoded representations of the two modalities:
\begin{equation}
    S = [X_{A, \text{enc}}, X_{B, \text{enc}}] \in \mathbb{R}^{2 \times D}
\end{equation}
This sequence $S$ is passed through a second BiMamba block. The recurrent nature of the SSM allows the hidden state generated by the first modality to dynamically influence the processing of the second modality, effectively modeling the biological ``crosstalk'' between the entities:
\begin{equation}
    S_{\text{mixed}} = \text{BiMambaBlock}(S)
\end{equation}
The final interaction vector $V_{\text{inter}}$ is obtained via global mean pooling across the sequence dimension.

\subsection{Prediction Head and Optimization}
The fused representation $V_{\text{inter}}$ is passed through a multi-layer perceptron (MLP) to generate the final prediction. For binary interaction tasks (RNA-Protein, RNA-RNA), we identify class imbalance as a critical challenge. Standard cross-entropy loss often biases models toward the majority class. To address this, we employ the Focal Loss \cite{lin2017focal}:
\begin{equation}
    \mathcal{L}_{\text{Focal}} = -\alpha_t (1 - p_t)^\gamma \log(p_t)
\end{equation}
where $p$ is the predicted probability for the positive class, and:
\begin{equation}
    p_t = \begin{cases} p & \text{if } y = 1 \\ 1 - p & \text{otherwise} \end{cases}, \quad
    \alpha_t = \begin{cases} \alpha & \text{if } y = 1 \\ 1 - \alpha & \text{otherwise} \end{cases}
\end{equation}
The modulating factor $(1-p_t)^\gamma$ down-weights well-classified examples, focusing training on hard negatives, while $\alpha_t$ provides additional class balancing. We set $\gamma=2.0$ and $\alpha=0.25$, thereby down-weighting easy examples and focusing training on hard-to-classify interactions to result in higher specificity and recall.
For RNA-Small Molecule binding affinity prediction, we optimize a composite loss function combining Mean Squared Error (MSE) and Pearson Correlation constraints to ensure the predicted affinity values correlate linearly with experimental ground truth.

\begin{algorithm}[t]
\caption{CrossLLM-Mamba Pipeline}
\label{alg:crossmamba}
\begin{algorithmic}[1]
\REQUIRE Biological Sequences $S_A, S_B$; Pretrained LLMs $\mathcal{M}_A, \mathcal{M}_B$
\REQUIRE Hyperparameters: Hidden dim $D$, Noise scale $\sigma$
\ENSURE Interaction Score $\hat{y} \in [0, 1]$ or Affinity $K_d \in \mathbb{R}$

\STATE \textbf{Stage 1: Multi-Modal Feature Extraction}
\STATE $E_A \leftarrow \mathcal{M}_A(S_A)$ \COMMENT{e.g., ESM-2 for Protein}
\STATE $E_B \leftarrow \mathcal{M}_B(S_B)$ \COMMENT{e.g., RiNALMo for RNA}

\STATE \textbf{Stage 2: Alignment \& Robustness}
\FOR{$k \in \{A, B\}$}
    \STATE $X_k \leftarrow \text{Linear}_{k \to D}(E_k) + \mathcal{N}(0, \sigma^2)$ \COMMENT{Project \& Inject Noise}
\ENDFOR

\STATE \textbf{Stage 3: Bidirectional Encoding (BiMamba)}
\FOR{$Z \in \{X_A, X_B\}$}
    \STATE $H_{\text{fwd}} \leftarrow \text{Mamba}(Z)$; \quad $H_{\text{bwd}} \leftarrow \text{Flip}(\text{Mamba}(\text{Flip}(Z)))$
    \STATE $Z \leftarrow \text{LayerNorm}(\text{Linear}([H_{\text{fwd}} \mathbin{\|} H_{\text{bwd}}]) + Z)$
\ENDFOR

\STATE \textbf{Stage 4: Cross-Mamba Fusion \& Prediction}
\STATE $S \leftarrow [X_A, X_B]$ \COMMENT{Stack aligned features}
\STATE $V_{\text{inter}} \leftarrow \text{MeanPool}(\text{BiMambaBlock}(S))$
\STATE $\hat{y} \leftarrow \text{MLP}(V_{\text{inter}})$

\IF{Task is Classification} \STATE $\hat{y} \leftarrow \sigma(\hat{y})$ \ENDIF
\RETURN $\hat{y}$
\end{algorithmic}
\end{algorithm}

\section{Experimental Setup}
\subsection{Datasets}
To rigorously evaluate the generalizability of CrossLLM-Mamba, we conducted experiments on three distinct benchmarks covering RNA-Protein, RNA-Small Molecule, and RNA-RNA interactions.

\subsubsection{RNA-Protein Interactions}
For the RNA-protein interaction (RPI) task, we utilize the RPI1460 benchmark dataset, which was originally introduced in the IPMiner study \cite{pan2016ipminer} and has been widely adopted in subsequent literature, including the LPI-CSFFR framework \cite{huang2022lpi}. This dataset consists of 1,460 experimentally validated interacting pairs between long non-coding RNAs (lncRNAs) and proteins, alongside an equal number of non-interacting pairs generated via random sampling to maintain a balanced distribution for binary classification. Following standard protocols, we employed 5-fold cross-validation to ensure robust performance estimation.

\begin{table*}[t]
\centering
\caption{Performance comparison on the RPI1460 dataset. Best values are in bold.}
\resizebox{\textwidth}{!}{
\begin{tabular}{lcccccc}
\toprule
\textbf{Method} & \textbf{MCC} & \textbf{ACC} & \textbf{F1} & \textbf{Precision} & \textbf{Recall} & \textbf{AUC--ROC} \\
\midrule
RPISeq-RF\cite{muppirala2011predicting} & 0.570 & 0.780 & 0.780 & 0.790 & 0.780 & 0.790 \\
IPMIner \cite{pan2016ipminer, pan2019recent}& 0.520 & 0.760 & 0.770 & 0.720 & 0.830 & 0.801 \\
CFRP \cite{dai2019construction} & 0.630 & 0.810 & 0.820 & 0.830 & 0.780 & 0.834 \\
RPITER\cite{peng2019rpiter} & 0.412 & 0.690 & 0.510 & 0.610 & 0.480 & 0.720 \\
LPI-CSFFR \cite{huang2022lpi} & 0.600 & 0.830 & 0.840 & 0.780 & 0.910 & 0.820 \\
RNAincoder \cite{wang2023rnaincoder} & 0.760 & 0.880 & 0.840 & 0.810 & 0.940 & 0.915 \\
BioLLMNet \cite{abir2025biollmnet} & 0.848 & 0.923 & 0.925 & 0.888 & 0.966 & 0.948 \\
\midrule
\textbf{CrossLLM-Mamba (Ours)} 
& \textbf{0.892} 
& \textbf{0.935 } 
& \textbf{0.933 } 
& \textbf{0.901 } 
& \textbf{0.971 } 
& \textbf{0.957} \\
\bottomrule
\end{tabular}
}
\label{tab:rpi1460_comparison}
\end{table*}

\subsubsection{RNA-Small Molecule Interactions}
For binding affinity prediction, we utilized the dataset from RSAPred\cite{krishnan2024reliable}, which aggregates experimental binding data from the ROBIN repository. The dataset is categorized into six biologically relevant RNA subtypes: \textit{Aptamers}, \textit{miRNAs}, \textit{Repeats}, \textit{Ribosomal RNAs}, \textit{Riboswitches}, and \textit{Viral RNAs}. We treated this as a regression task to predict binding affinity values and performed 10-fold cross-validation. Additionally, for binary classification metrics, we applied a threshold (affinity $\geq 4.0$) to categorize interactions as active or inactive, consistent with baseline methodologies.

\subsubsection{RNA-RNA Interactions}
To assess cross-species generalization, we employed three plant-based miRNA-lncRNA interaction datasets collected from \textit{Arabidopsis thaliana} (Ath), \textit{Glycine max} (Gma), and \textit{Medicago truncatula} (Mtr). Following the CORAIN benchmark protocol \cite{wang2023corain}, we evaluated the model using a ``train-on-one, test-on-another'' strategy, resulting in six distinct transfer learning scenarios to measure the model's robustness to unseen biological contexts.

\subsection{Implementation Details}
The proposed framework was implemented using PyTorch and the Mamba-SSM library. All experiments were conducted on NVIDIA GPUs. Across all tasks, we projected modality-specific embeddings into a shared latent dimension of $D=512$. The BiMamba encoders were configured with a state dimension of 8 or 16, convolution kernel size of 4 or 8, and an expansion factor of 2 or 4 depending on the task complexity.

\begin{table*}[t]
\centering
\caption{Comparison of BioLLMNet \cite{abir2025biollmnet}, RLaffinity \cite{sun2024rlaffinity}, RSAPred \cite{krishnan2024reliable}, and CrossLLM-Mamba. Best values are in bold.}
\resizebox{\textwidth}{!}{
\begin{tabular}{l l c c c c c c}
\toprule
\textbf{Metric} & \textbf{Method} 
& \textbf{Aptamer} 
& \textbf{Repeats} 
& \textbf{Ribosomal} 
& \textbf{Riboswitch} 
& \textbf{Viral} 
& \textbf{miRNA} \\
\midrule

\multirow{3}{*}{\textbf{Pearson}}
& BioLLMNet 
& 0.7713 
& 0.9340 
& 0.8933 
& 0.9462 
& 0.8004 
& 0.8712 \\
& RLaffinity 
& 0.7400 & 0.9100 & 0.8500 & 0.9250 & 0.7890 & 0.8750 \\
& RSAPred 
& 0.7180 & 0.8940 & 0.8360 & 0.9090 & 0.7840 & \textbf{0.8810}\\
\midrule

& \textbf{CrossLLM-Mamba (Ours)}
& \textbf{0.8254} 
& \textbf{0.9521} 
& \textbf{0.9012 } 
& \textbf{0.9562 } 
& \textbf{0.829} 
& 0.869  \\
\midrule

\multirow{3}{*}{\textbf{MAE}}
& BioLLMNet 
& 0.5726 
& 0.3572 
& 0.5743 
& 0.5236 
& 0.5582 
& \textbf{0.3827} \\
& RLaffinity 
& 0.5900 & 0.3630 & 0.6400 & 0.5300 & 0.6010 & 0.4000 \\
& RSAPred 
& 0.6070 & 0.3670 & 0.6870 & 0.5370 & 0.6070 & 0.4340 \\
\midrule

& \textbf{CrossLLM-Mamba (Ours)}
& \textbf{0.5315 } 
& \textbf{0.3410 } 
& \textbf{0.5621 } 
& \textbf{0.5020} 
& \textbf{0.5112} 
& 0.3914 \\
\bottomrule
\end{tabular}
}
\label{tab:affinity_comparison}
\end{table*}

\begin{table*}[t]
\centering
\caption{Comparison of CrossLLM-Mamba on miRNA--lncRNA interaction datasets based on accuracy (\%), 
where the best values are shown in bold.}
\resizebox{\textwidth}{!}{
\begin{tabular}{lcccccc}
\toprule
\textbf{Model} & \textbf{ATH-GMA} & \textbf{ATH-MTR} & \textbf{GMA-ATH} 
& \textbf{GMA-MTR} & \textbf{MTR-ATH} & \textbf{MTR-GMA} \\
\midrule
PMLIPred \cite{kang2020pmlipred} & 65 & 70 & 55 & 87 & 53 & 71 \\
CORAIN \cite{wang2023corain} & 69 & 74 & 67 & \textbf{93} & 58 & 84 \\
BioLLMNet \cite{abir2025biollmnet} & 69 
& 72  
&  69 
& 84  
& 68 
& 85 \\
\midrule
\textbf{CrossLLM-Mamba (Ours)} 
& \textbf{72 } 
& \textbf{74} 
& \textbf{73 } 
& 87 
& \textbf{75 } 
& \textbf{85} \\
\bottomrule
\end{tabular}
}
\label{tab:rna-rna comparison}
\end{table*}


\subsection{Baseline Models}
To evaluate the performance of CrossLLM-Mamba, we compare it against a range of baselines including traditional machine learning, deep learning architectures, and recent LLM-based frameworks. These models are selected based on their established performance on the benchmarks used in this study.

\subsubsection{RNA-Protein Interaction (RPI) Baselines}
The models evaluated on the RPI1460 benchmark include:
\begin{itemize}
    \item \textbf{RPISeq-RF} \cite{muppirala2011predicting}: A sequence-based method utilizing Random Forest with 3-mer and 4-mer features.
    \item \textbf{IPMiner} \cite{pan2016ipminer}: A foundational framework utilizing stacked autoencoders to extract features from RNA and protein sequences.
    \item \textbf{CFRP} \cite{dai2019construction}: A model focused on the construction of complex features for ncRNA-protein interaction prediction.
    \item \textbf{RPITER} \cite{peng2019rpiter}: A hierarchical deep learning framework specifically designed for lncRNA-protein pairs.
    \item \textbf{LPI-CSFFR} \cite{huang2022lpi}: A framework that employs serial fusion and feature reuse for interaction prediction.
    \item \textbf{RNAincoder} \cite{wang2023rnaincoder}: A state-of-the-art encoder based on convolutional autoencoders for RNA-associated interactions.
    \item \textbf{BioLLMNet} \cite{abir2025biollmnet}: A contemporary framework utilizing cross-LLM transformation networks for embedding fusion.
\end{itemize}

\subsubsection{RNA-Small Molecule and RNA-RNA Baselines}
For binding affinity and cross-species transfer tasks, we compare against:
\begin{itemize}
    \item \textbf{RSAPred} \cite{krishnan2024reliable}: A machine learning-based method for RNA-small molecule affinity prediction.
    \item \textbf{RLaffinity} \cite{sun2024rlaffinity}: A method combining contrastive pre-training with 3D CNNs for affinity estimation.
    \item \textbf{PmliPred} \cite{kang2020pmlipred}: A hybrid model designed for plant miRNA-lncRNA interaction prediction.
    \item \textbf{CORAIN} \cite{wang2023corain}: A task-specific encoding algorithm for cross-species RNA interaction generalization.
    \item \textbf{BioLLMNet} \cite{abir2025biollmnet}: Serving as a unified baseline across all three interaction categories.
\end{itemize}

\subsection{Evaluation Metrics}
To comprehensively assess the performance of CrossLLM-Mamba across different tasks, we utilize a diverse set of evaluation metrics tailored to specific interaction types.
For binary interaction prediction tasks, we report Accuracy (ACC), Precision, Recall (Sensitivity), F1-score, and the Area Under the Receiver Operating Characteristic Curve (AUC-ROC). Additionally, given the prevalent class imbalance in biological interaction datasets, we rely heavily on the Matthews Correlation Coefficient (MCC), which is regarded as a more robust metric for imbalanced classes than accuracy or F1-score. The MCC is calculated as:
\begin{equation}
    \text{MCC} = \frac{(TP \times TN) - (FP \times FN)}{\sqrt{(TP+FP)(TP+FN)(TN+FP)(TN+FN)}}
\end{equation}
where $TP$, $TN$, $FP$, and $FN$ represent true positives, true negatives, false positives, and false negatives, respectively.

For binding affinity prediction, we evaluate the linear correlation and error magnitude between the predicted affinity scores and experimental values. We utilize the Pearson Correlation Coefficient (PCC) to measure the linear relationship:
\begin{equation}
    \text{PCC} = \frac{\sum (y_i - \bar{y})(\hat{y}_i - \bar{\hat{y}})}{\sqrt{\sum (y_i - \bar{y})^2 \sum (\hat{y}_i - \bar{\hat{y}})^2}}
\end{equation}
Additionally, we report the Mean Absolute Error (MAE) to quantify the average magnitude of prediction errors.

\begin{figure}[htbp]
    \centering
    \includegraphics[width=\linewidth]{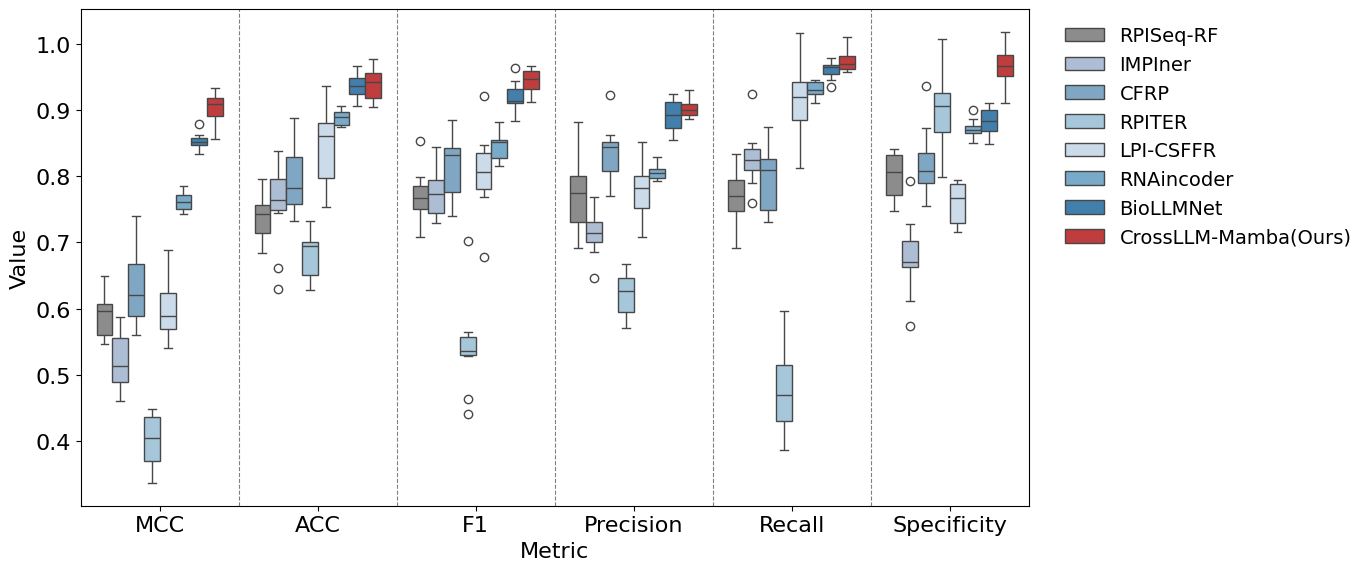}
    \caption{\textbf{Performance Comparison on the RPI1460 Dataset.} 
    The boxplots illustrate the distribution of performance metrics (MCC, ACC, F1, Precision, Recall, and Specificity) across 5-fold cross-validation for various state-of-the-art methods. 
    Our proposed \textbf{CrossLLM-Mamba} (shown in red) consistently outperforms existing baselines, achieving the highest median scores and lowest variance across all major metrics, particularly in MCC and F1-score, demonstrating its robustness and superior predictive capability.}
    \label{fig:method_comparison}
\end{figure}
\subsection{Result Analysis}
The experimental results across three heterogeneous biological interaction tasks demonstrate the robust predictive power and generalization capabilities of CrossLLM-Mamba.

\subsubsection{Superiority in RNA-Protein Interaction Prediction}
As shown in Table \ref{tab:rpi1460_comparison} and Fig. \ref{fig:method_comparison}, CrossLLM-Mamba achieves state-of-the-art performance on the RPI1460 benchmark. Our model reaches an MCC of 0.892 and an accuracy of 0.935, representing a significant improvement over the previous best-performing model. 

The high recall value ($0.971$) is particularly noteworthy; it suggests that the state-space alignment mechanism is effective at identifying true positive interactions within the high-dimensional latent space of ESM-2 and RiNALMo. The reduced variance shown in the boxplots (Fig. \ref{fig:method_comparison}) further confirms that the integration of Gaussian noise injection and BiMamba encoding leads to more stable learning across cross-validation folds compared to static architectures like RPISeq or CFRP.

\subsubsection{Precision in RNA-Small Molecule Binding Affinity}
The results for binding affinity regression (Table \ref{tab:affinity_comparison}) highlight the model's ability to capture fine-grained chemical dependencies. CrossLLM-Mamba outperforms RSAPred and RLaffinity across nearly all RNA subtypes. 
\begin{itemize}
    \item \textbf{Linear Correlation:} The model achieves a Pearson correlation of $0.9562$ for Riboswitches and $0.9521$ for Repeats, suggesting that the Cross-Mamba Interaction Module effectively maps the ``crosstalk'' between RNA sequences and MoleBERT molecular graphs.
    \item \textbf{Error Reduction:} We observe a consistent decrease in Mean Absolute Error (MAE) across most categories, notably in the Viral RNA class. 
\end{itemize}
While the miRNA category showed a slightly higher PCC in RSAPred, CrossLLM-Mamba maintains highly competitive performance, demonstrating that a unified SSM-based architecture can handle both binary classification and continuous regression tasks with minimal task-specific tuning.

\subsubsection{Generalization in RNA-RNA Cross-Species Transfer}
The plant-based miRNA-lncRNA interaction tasks (Table \ref{tab:rna-rna comparison}) provide a rigorous test of the model's ability to generalize to unseen biological distributions. CrossLLM-Mamba outperforms the CORAIN and BioLLMNet baselines in four out of six transfer learning scenarios. 
Specifically, in the \textbf{MTR-ATH} scenario (training on \textit{M. truncatula} and testing on \textit{A. thaliana}), our model achieves a $75\%$ accuracy, a substantial $7\%$ improvement over BioLLMNet. This suggests that the bidirectional state-space modeling captures universal structural motifs in RNA that are conserved across species, which are often missed by models that rely solely on sequence-level concatenation. The slightly lower performance in the GMA-MTR task compared to CORAIN indicates that while Mamba is highly efficient, sequence-specific motifs in certain species pairs may still benefit from localized attention mechanisms, which we intend to explore in future work.

\section{Ablation Study}
\label{sec:ablation}

To validate the architectural choices of CrossLLM-Mamba, we performed a component-wise ablation study on the RNA-Protein interaction benchmark (RPI1460). We systematically removed or replaced key modules specifically the Cross-Mamba fusion mechanism, the bidirectional scanning, and the focal loss to quantify their individual contributions to predictive performance. 

\subsection{Impact of Interaction Fusion Mechanism}
\begin{figure}[h]
    \centering
    \includegraphics[width=\linewidth]{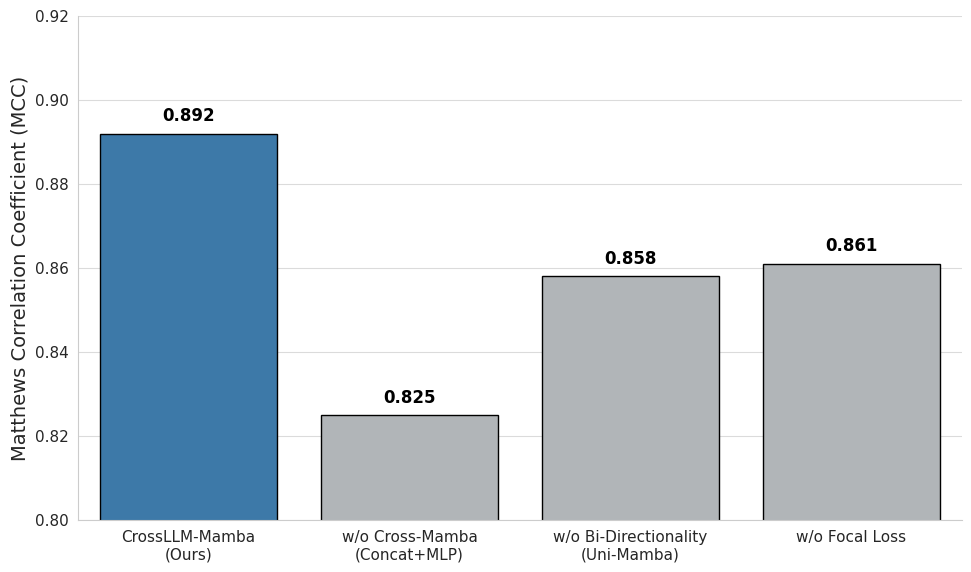} 
    \caption{Performance impact of removing specific architectural components. The full CrossMamba-Bio model (blue) significantly outperforms variants lacking the cross-modal state mixing or bidirectional context.}
    \label{fig:ablation}
\end{figure}

A core hypothesis of this work is that interaction is a dynamic state transition rather than a static feature overlap. To test this, we replaced our Cross-Mamba Interaction Module with a standard Concatenation and MLP baseline. As shown in Fig. \ref{fig:ablation}, replacing the Cross-Mamba module with simple concatenation resulted in the most significant performance drop. This confirms that the state-space mixing (SSM) approach more effectively captures the continuous information flow between biological modalities compared to static feature aggregation.

\subsection{Importance of Bidirectionality in Embeddings}
Biological sequences, unlike temporal sentences, possess spatial structures where "future" tokens (downstream sequence) influence "past" tokens (upstream sequence) via folding. We compared our \textbf{BiMamba} encoder against a standard unidirectional Mamba encoder. The unidirectional variant suffered a performance degradation of $2.7\%$ in MCC. This confirms that scanning biological embeddings in both forward and reverse directions is essential to capture the non-causal structural dependencies inherent in protein and RNA tertiary structures.

\subsection{Robustness via Noise Injection}
Finally, we analyzed the contribution of our Gaussian Noise Injection strategy ($\sigma=0.02$). Removing this regularization led to a higher training accuracy but a lower validation F1-score, indicating overfitting. The noise injection forces the model to learn smoother decision boundaries in the high-dimensional LLM latent space, proving critical for generalizing to unseen sequences.

\subsection{Analysis of Components}
\begin{itemize}
    \item \textbf{Significance of Cross-Mamba Fusion:} The most substantial drop in performance occurred when replacing our Cross-Mamba interaction module with a standard concatenation and MLP head. This confirms our hypothesis that modeling interaction as a dynamic state transition is superior to static feature aggregation.
    
    \item \textbf{Necessity of Bidirectionality:} Reverting to a standard unidirectional Mamba encoder resulted in a $2.7\%$ drop in MCC. Since biological embeddings represent global molecular structures rather than temporal sequences, the bidirectional scan is essential for capturing non-causal dependencies.
    
    \item \textbf{Robustness Features:} The removal of Gaussian noise injection and Focal loss resulted in smaller but significant drops. Notably, removing Focal loss degraded specificity, indicating that the model struggled to distinguish hard-negative samples without the focused training objective.
\end{itemize}
\subsection{Ablation on Encoder and Fusion BiMamba Depth.}
\begin{figure}[h]
    \centering
    \includegraphics[width=\linewidth]{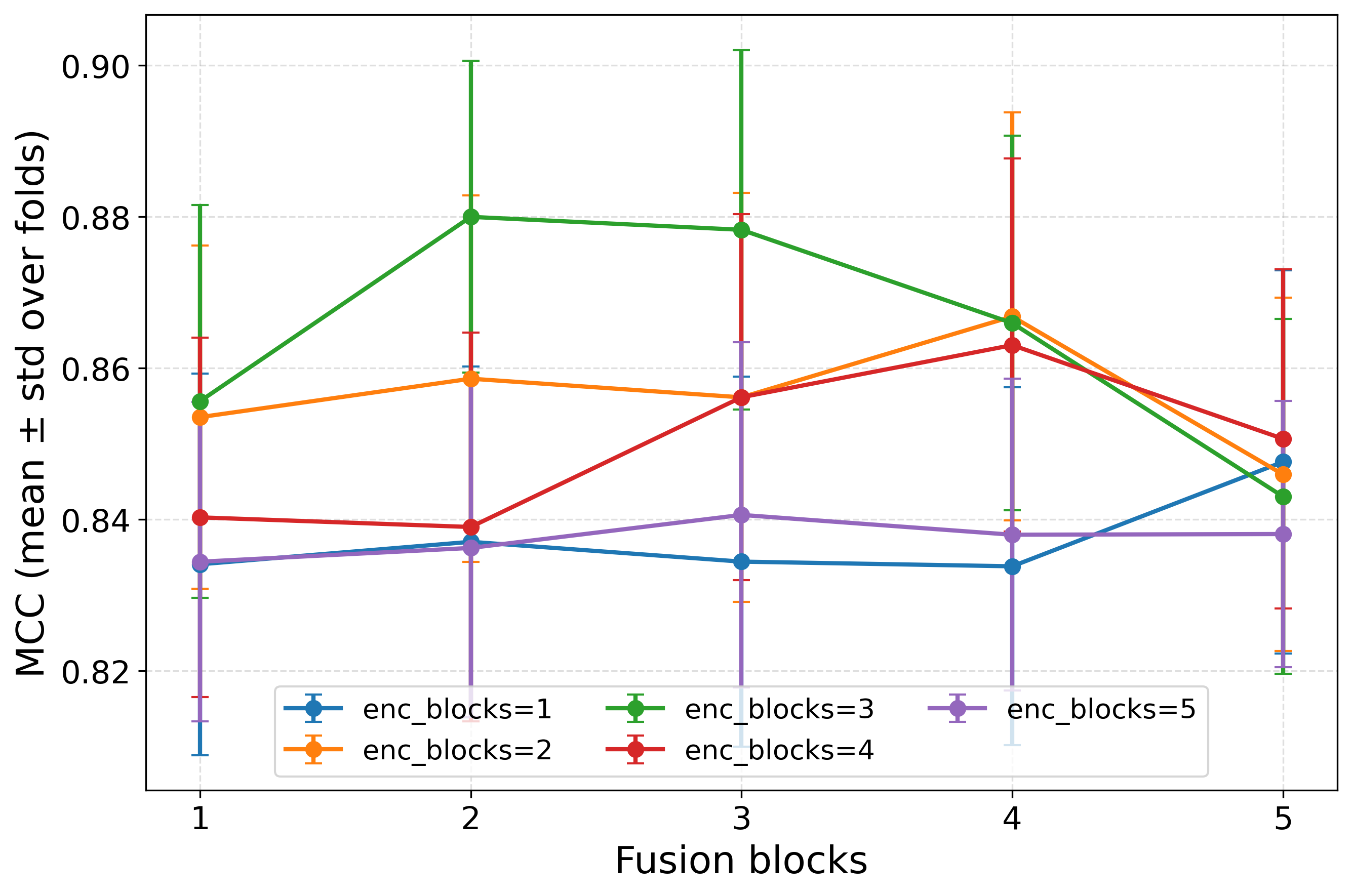} 
    \caption{Ablation study on the number of BiMamba blocks in the modality-specific encoders and fusion module. Performance peaks at moderate depth (\texttt{enc\_blocks}=3, \texttt{Fusion blocks}=2--3), while deeper stacks show diminishing returns.}
    \label{fig:ablation_bimamba_blocks}
\end{figure}

We investigate the impact of the number of BiMamba blocks used in the modality-specific encoders (\texttt{enc\_blocks}) and the cross-modal fusion module (\texttt{Fusion blocks}). 
Figure~\ref{fig:ablation_bimamba_blocks} reports the MCC for different encoder--fusion depth combinations. 
Overall, we observe that increasing \texttt{Fusion blocks} improves performance when the encoder depth is moderate, with the best results achieved at \texttt{enc\_blocks}=3 and \texttt{Fusion blocks}=2--3. However, further increasing fusion depth beyond this point leads to diminishing returns and even performance degradation, suggesting that overly deep fusion can over-mix the two modalities and reduce discriminative interaction signals. 
Similarly, very shallow encoders (\texttt{enc\_blocks}=1) underperform due to insufficient representation refinement, while very deep encoders (\texttt{enc\_blocks}=5) consistently yield lower MCC, indicating potential over-refinement before fusion. 
These results confirm that moderate encoder refinement combined with moderate cross-modal interaction modeling provides the most effective and stable configuration.

\section{Conclusion}

In this work, we introduce CrossLLM-Mamba, a novel framework that reconceptualizes biological interaction prediction as a state-space alignment problem. By leveraging the efficiency of selective state-space models, our approach enables dynamic ``crosstalk'' between embeddings from specialized Biological Large Language Models ESM-2 for proteins, RiNALMo for RNA, and MoleBERT for small molecules through hidden state propagation in a bidirectional Mamba architecture. Unlike conventional static fusion strategies that treat interaction as mere feature overlap, our Cross-Mamba Interaction Module models the mutual conditioning between molecular partners while maintaining linear computational complexity.

Our comprehensive experimental evaluation across three distinct interaction categories demonstrates the broad applicability of this paradigm. For RNA-protein interaction prediction, CrossLLM-Mamba achieved state-of-the-art performance with an MCC of 0.892 and recall of 0.971 on the RPI1460 benchmark. In RNA-small molecule binding affinity prediction, the model attained Pearson correlations of 0.956 for riboswitches and 0.952 for repeats, outperforming existing methods across most RNA subtypes. The cross-species transfer experiments on plant miRNA-lncRNA interactions further validated the model's generalization capability, with substantial improvements in challenging scenarios such as MTR-ATH transfer.

\textbf{Limitations.} While CrossLLM-Mamba demonstrates strong performance, several limitations should be acknowledged. First, the model operates on sequence-level embeddings and does not explicitly incorporate three-dimensional structural information, which may limit performance for interactions dominated by specific binding motifs. Second, the slightly lower performance in certain cross-species scenarios (e.g., GMA-MTR) suggests that localized attention mechanisms may complement our global state-space approach for capturing species-specific sequence patterns. Third, our current framework predicts interaction likelihood but does not identify specific binding residues or sites.

\textbf{Future Directions.} Promising extensions include integrating predicted or experimental 3D structural features, developing hybrid architectures that combine Mamba with sparse attention for local motif detection, and extending the framework to predict residue-level interaction sites. The success of state-space modeling demonstrated here opens new avenues for applying SSM-based architectures to other multi-modal problems in computational biology, including drug-target interaction and protein-protein interaction prediction.

\section*{Acknowledgements}
This research is supported in part by the NSF under Grant IIS 2327113 and ITE 2433190, the NIH under Grants R21AG070909 and P30AG072946, and the National Artificial Intelligence Research Resource (NAIRR) Pilot NSF OAC 240219, and Jetstream2, Bridges2, and Neocortex resources. We thank the University of Kentucky Center for Computational Sciences and Information Technology Services Research Computing for their support and use of the Lipscomb Compute Cluster and associated research computing resources.  
\bibliographystyle{elsarticle-num}
\bibliography{sample-base}

\end{document}